\documentclass[aps,prd,preprint,groupedaddress]{revtex4-1}

\usepackage{amsfonts,amssymb,amsmath}
\usepackage{hyperref,graphicx}
\usepackage[dvipsnames]{xcolor}
\newcommand{\rb}{\mbox{\boldmath{$b$}}}
\newcommand{\rd}{\mbox{\boldmath{$\Delta$}}}
\newcommand{\rr}{\mbox{\boldmath{$r$}}}
\newcommand{\dfr}{\mathrm{d}}

\begin{document}


\title{Effect of sub-nucleon fluctuations on the DVCS process in proton and nuclear targets at the EIC}


\author{J.~Cepila}
\email{jan.cepila@fjfi.cvut.cz}
\affiliation{Faculty of Nuclear Sciences and Physical Engineering, Czech Technical University in Prague, Czech Republic}

\author{V.~P.~Gon\c calves}
\email{barros@ufpel.edu.br}
\affiliation{Institute of Physics and Mathematics, Federal University of Pelotas (UFPel), \\
  Postal Code 354,  96010-900, Pelotas, RS, Brazil}

\author{A.~Ridzikova}
\email{ridziale@fjfi.cvut.cz}
\affiliation{Faculty of Nuclear Sciences and Physical Engineering, Czech Technical University in Prague, Czech Republic}


\date{\today}
\begin{abstract}
The impact of the sub - nucleon fluctuations on the Deeply Virtual Compton Scattering (DVCS) process at the Electron - Ion Collider (EIC)  is investigated considering proton and nuclear targets. Assuming the hot - spot model, we estimate the energy dependence of the  coherent and incoherent cross - sections for different values of the photon virtuality and atomic number. Predictions for the $t$ - distributions are also presented. We demonstrate that the sub - nucleon fluctuations in the proton, as described by the hot - spot model, implies a turn - over in the energy dependence of the incoherent cross - section, with the position of the maximum being dependent of the photon virtuality. Our results indicate that the ratio between the coherent and incoherent cross - sections increases with energy, atomic numbers and for smaller values of $Q^2$. Moreover, we predict a maximum in the $t$ - distribution of the nuclear incoherent cross - section at a fixed center - of - mass energy, which is dependent on the atomic number and $Q^2$. 

\end{abstract}


\maketitle

\section{Introduction}
The deeply virtual Compton scattering (DVCS) process  is a diffractive process, where the electron scatters off a hadron by exchanging a virtual photon and a real photon is produced in the final state.  The interaction is mediated by a color - singlet object and can be classified as coherent, if the hadron target remains intact, or incoherent if it dissociates  into an inclusive multi-hadron state $Y$. In both cases, the real photon and the hadronic system are  separated by a large rapidity gap. In recent years, several studies have demonstrated that the DVCS process is one of the most promising ways for probing the hadronic structure \cite{McDermott:2001pt, Favart:2003cu, Kowalski:2006hc, Machado:2008tp, Aschenauer:2013hhw, Goncalves:2015goa, Hatta:2017cte, Mantysaari:2020lhf, Bendova:2022xhw, Goncalves:2022wzq, Xie:2022sjm, Xiang:2024fee}.  As this process is driven at high energies by the gluon 
content of the target, with the  cross - section being  proportional to the square of the scattering amplitude, its description is strongly sensitive to the underlying QCD dynamics (See, e.g. Refs. \cite{Goncalves:2015goa,Bendova:2022xhw}).
In addition, the experimental analysis of the differential cross - sections $d\sigma/dt$, where $t$ is the squared four - momentum transfer between the incoming and scattered hadron, is expected to allow the determination of the transverse spatial distributions of quarks and gluons in both protons and  nuclei\cite{Klein:2019qfb}. In particular, in the Good - Walker  approach \cite{Good:1960ba},  the coherent cross - section  measures the average spatial distribution of gluons in the target, while the incoherent one  the fluctuations and correlations in the gluon density \cite{Mantysaari:2020axf}.

In this paper, we will investigate the impact of sub - nucleon fluctuations on the coherent and incoherent DVCS cross - sections considering the kinematical range that will be probed by the  the electron - ion collider (EIC) at the Brookhaven National Laboratory \cite{eic}.
In our analysis, we will consider the model proposed in Ref.\cite{Cepila:2016uku}, in which the quantum fluctuation of the proton are characterized by hot - spots, whose number increases with energy and  have positions in the impact - parameter plane that change event - by - event. Such a model was successfully applied for the vector - meson photoproduction at HERA and LHC in Refs.~\cite{Cepila:2016uku,Cepila:2017nef,Cepila:2018zky,Cepila:2023dxn,Cepila:2025exl} and extended for the EIC in Refs.~\cite{Bendova:2018bbb,Krelina:2019gee}. In these references,  the authors also have demonstrated that the energy dependence on the number of hot - spots implies that the incoherent cross - section for a proton target reaches a maximum after which it decreases steeply at higher energies, with the position of the maximum being dependent of the mass of vector meson. Our goal in this paper is to extend these analyze for the DVCS process, and to present, for the first time, the associated predictions for the coherent and incoherent cross - sections. In particular, we will investigate the energy, photon virtuality, transferred momentum and atomic number dependencies of these cross - sections and present predictions at the EIC energies.

This paper is organized as follows. In the next section, we will present a brief review of the Good - Walker and color - dipole approaches, and present the main ingredients assumed in our calculations. In section \ref{sec:results},  we  present our predictions for the energy, $t$, $Q^2$ and atomic dependencies of the coherent and incoherent cross - sections considering the expected EIC energies. In addition, we  present the results for the energy dependence of the ratio between the coherent and incoherent cross - sections. A comparison of our predictions with the HERA data for the DVCS process is also performed, which demonstrate that the hot - spot model is able to describe the current data. Finally, in section \ref{sec:summary}, we summarize our main results and conclusions.

\section{Formalism}
In order to estimate the cross - sections associated with the   coherent and incoherent  DVCS processes, we will consider the Good - Walker approach \cite{Good:1960ba}. In this approach, the differential cross - section, $d\sigma/dt$,  for the coherent process is expressed  as follows
\begin{equation}
\frac{\mathrm{d}\sigma^{\gamma^* h \rightarrow \gamma h }}{\mathrm{d}t}
 =  \frac{1}{16\pi} \bigg|\sum_f R^f_g \left<A_f(x, Q^2,  \rd)\right>\bigg|^2\,\,,
\label{dsigdtcoherent}
\end{equation}
while for an incoherent interaction is given by
\begin{equation}
\frac{\mathrm{d}\sigma^{\gamma^* h \rightarrow \gamma Y }}{\mathrm{d}t}
 =  \frac{1}{16\pi} \left( \bigg|\sum_f R^f_g \sqrt{\left<A_f^2(x, Q^2,  \rd)\right>}\bigg|^2  - 
\bigg|\sum_f R^f_g \left<A_f(x, Q^2,  \rd )\right>\bigg|^2 \right)\,.
\label{dsigdtincoherent}
\end{equation}
The sum runs over all flavors, $x$ is the Bjorken variable, $Q^2$ is the virtuality of the incoming photon and $|\rd| =  \sqrt{-t}$ is the momentum transfer. Moreover, the skewedness correction $R_g^f$ can be expressed as \cite{Kowalski:2006hc} 
\begin{equation}
R^f_g = \frac{2^{2\lambda^f_g+3}}{\sqrt{\pi}}\frac{\Gamma(\lambda^f_g+5/2)}{\Gamma(\lambda^f_g+4)}\qquad 
\lambda^f_g\equiv \Bigg|\frac{\partial\ln(A_f(x,Q^2,\rd=0))}{\partial\ln(1/x)}\Bigg|.
\end{equation}

In the color - dipole model \cite{Nikolaev:1990ja,Nikolaev:1991et}, the scattering amplitude for the $\gamma^* h  \rightarrow \gamma h$ process  can be factorized in terms of the fluctuation of the incoming virtual photon, emitted by the incident electron, in a 
$q\bar q$ Fock state, the dipole - hadron scattering by a color singlet exchange and the recombination into the exclusive photon final state. The final expression is given by \cite{Kowalski:2006hc}
\begin{equation}
 {\cal A}_f({x},Q^2,\rd)  =   i
\int \frac{\dfr z}{4\pi} \, \dfr^2\rr \, \dfr^2\rb_h \,  e^{-i\left[\rb_h+\left(\frac{1}{2}-z\right)\rr\right].\rd}  
 \,\, [\Psi_{\gamma} \Psi_{\gamma^*}]_T^f(z,\rr,Q^2)  \frac{\dfr\sigma^{\rm dip}}{\dfr \rb_h}(\tilde x,\rr),
\label{amp}
\end{equation}
where $\rb_h$ is the impact parameter of the dipole relative to the hadron target and the function $[\Psi_{\gamma} \Psi_{\gamma^*}]^f_T(z,\rr,Q^2)$ denotes the wave function overlap between the virtual photon of transverse polarization $T$ in the initial state fluctuating into $q\bar q$ pair with flavor $f$ and the real photon in the final state. Moreover, the variables  $\rr$ and $z$ are the dipole transverse radius and the momentum fraction of the photon momentum carried by a quark (an antiquark carries then $1-z$), respectively. Finally, $\dfr\sigma^{\rm dip}/\dfr \rb_h(\tilde x,\rr)$ is the forward dipole-target scattering amplitude (for a dipole at impact parameter $\rb_h$) with $\tilde{x} = x(1 + 4 m_f^2/Q^2)$ and $m_f$ being the mass of the quark.

The photon wave functions can be calculated via QED, and it is well known in literature \cite{Kowalski:2006hc}. In the DVCS case, as one has a real photon at the final state, only the transversely polarized overlap function contributes to the cross - section. Summed over the quark helicities, for a given quark flavour $f$ it is 
given by 
\begin{equation}
[\Psi_{\gamma} \Psi_{\gamma^*}]_T^f(z,\rr,Q^2) = \frac{2 N_c\,\alpha_{\mathrm{em}}e_f^2}{\pi}\left\{\left[z^2+\bar{z}^2\right]\varepsilon_1 K_1(\varepsilon_1 r)\varepsilon_2 K_1(\varepsilon_2 r) + m_f^2 K_0(\varepsilon_1 r) K_0(\varepsilon_2 r)\right\},
  \label{eq:overlap_dvcs}
\end{equation}
where $\varepsilon_{1,2}^2 = z\bar{z}\,Q_{1,2}^2+m_f^2$ and $\bar{z}=(1-z)$. Accordingly, the photon virtualities are $Q_1^2=Q^2$ (incoming virtual photon) and $Q_2^2=0$ (outgoing real photon).

The  dipole - hadron cross - section is determined by the QCD dynamics and by the transverse profile of the hadron target (see, e.g. Ref.~\cite{Morreale:2021pnn}). In our analysis, in order to estimate the impact of sub - nucleon fluctuations on the DVCS process, we will consider the  hot - spot model presented in Refs. \cite{Cepila:2017nef,Cepila:2018zky,Krelina:2019gee,Cepila:2023dxn,Cepila:2025exl}, which successfully describe the diffractive vector meson production at HERA. In this model, the cross - section of the interaction of the dipole with a proton is calculated assuming the factorization of the impact parameter dependence, as follows
\begin{equation}
\frac{\dfr\sigma^{\rm dip}}{\dfr\rb_p} = \sigma_0 N(\tilde{x},\rr)T(\rb_p),\qquad \sigma_0\equiv4\pi B_p\,\,,
\end{equation}
where $N(\tilde{x},\rr)$ is the forward scattering amplitude and $T(\rb_p)$ is the proton profile function. As in Refs. \cite{Cepila:2017nef,Cepila:2018zky,Krelina:2019gee,Cepila:2023dxn,Cepila:2025exl}, we will assume the GBW model \cite{Golec-Biernat:1998zce} to describe the forward scattering amplitude, which implies 
\begin{equation}
N(\tilde{x},\rr) = \left[ 1 - \exp \left( -\frac{\rr^2 Q_s ^2 (\tilde{x})}{4} \right) \right]
\label{dipole-proton}
\end{equation}
with
\begin{equation}
Q_s(x)= Q_0 ^2 \left( x_0/\tilde{x}\right) ^{\lambda}\qquad \lambda=0.21 \qquad x_0=0.0002.
\label{dipole-proton-param}
\end{equation}
 The transverse profile of a proton target is sampled with randomly distributed places of color charge (hot - spots) from the Gaussian distribution with a proton width $B_p=4.7 \text{ GeV}^{-2}$, being given by 
\begin{equation}
T_{\rm p}(\rb_{h}) = \frac{1}{2\pi B_{\rm hs}N_{\rm hs}} \sum \limits_{i=1}^{N_{\rm hs}} \exp \left( -\frac{\left( \rb_h - \rb_i\right)^2}{2B_{\rm hs}}\right)
\label{T_hs_proton}
\end{equation}
The width of the hot - spot is fixed to $B_{hs}=0.8 \text{ GeV}^{-2}$ and the number of hot - spots $N_{hs}$ is taken from the zero - truncated Poissonian distribution with energy - dependent mean 
\begin{equation}
\langle N_{ \mathrm{hs}}(x) \rangle = p_0x^{p_1}(1+p_2\sqrt{x}), \qquad p_0=0.025, \; p_1=-0.58, \; p_2= 300, 
\label{Nhsx}
\end{equation}
where $x$ corresponds to the Bjorken-$x$ variable. 

\begin{figure}[t]
\centering
\includegraphics[width=0.42\textwidth]{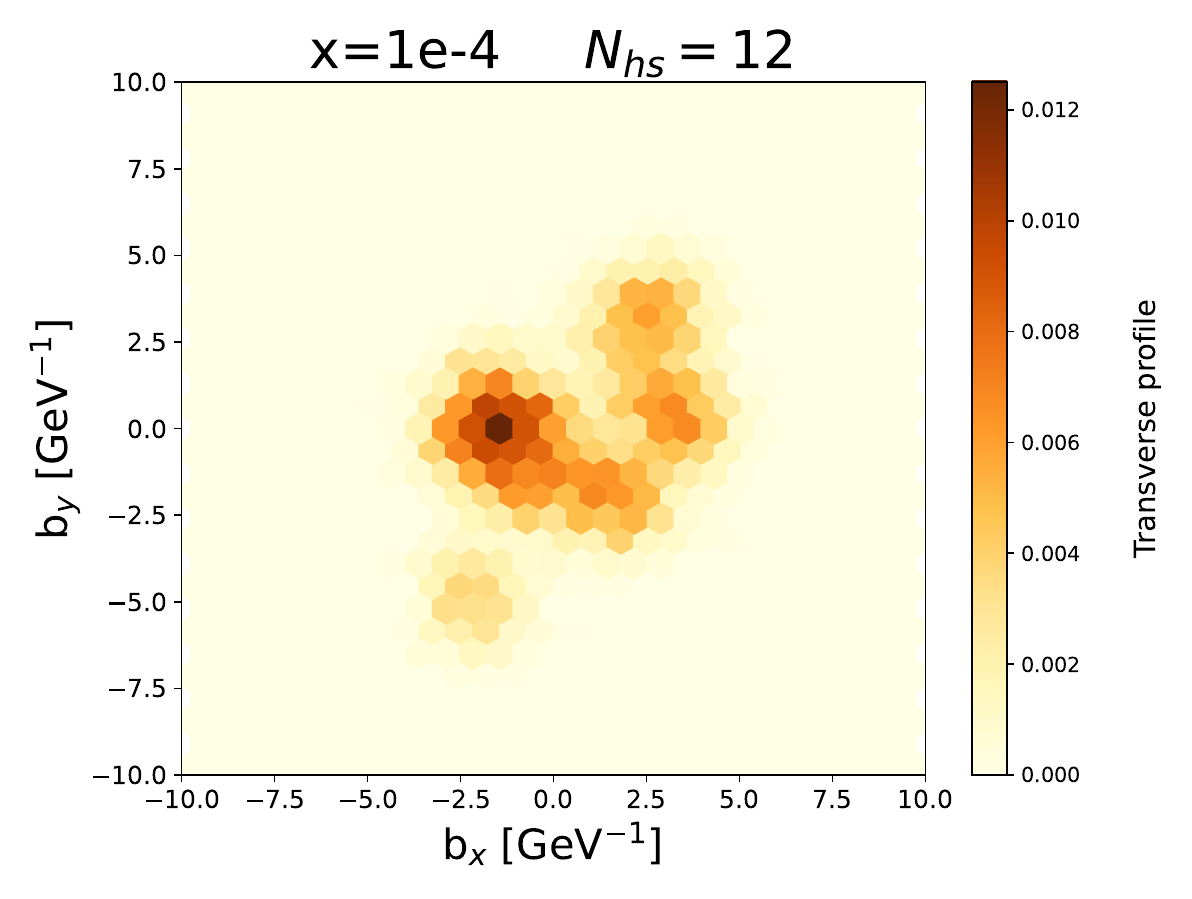}
\includegraphics[width=0.42\textwidth]{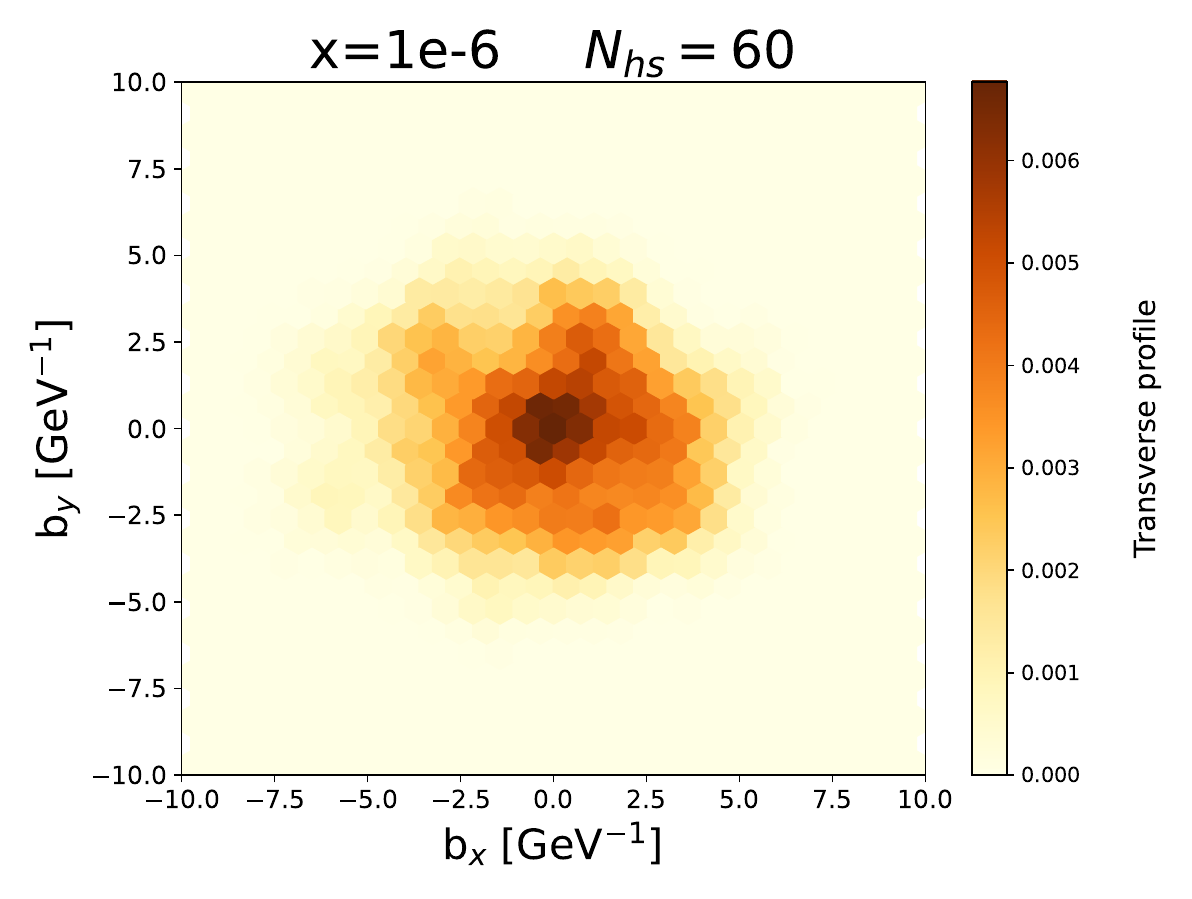}
 \centering
 \caption{Transverse profile of a proton generated using the energy - dependent hot - spot model at $ x=10^{-4}$ (left panel) and $ x=10^{-6}$ (right panel).}
 \label{fig:HS_proton}
\end{figure}

\begin{figure}[t]
\centering
\includegraphics[width=0.42\textwidth]{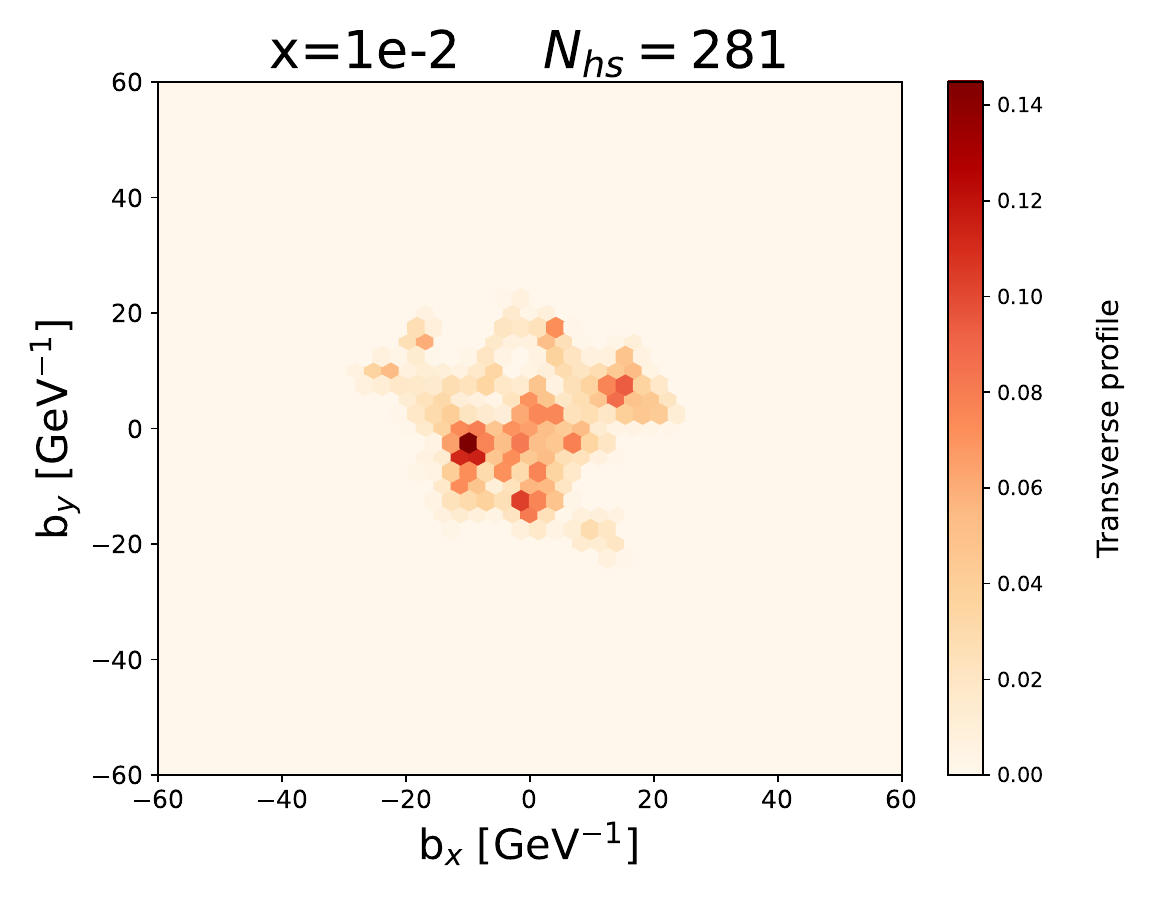}
\includegraphics[width=0.42\textwidth]{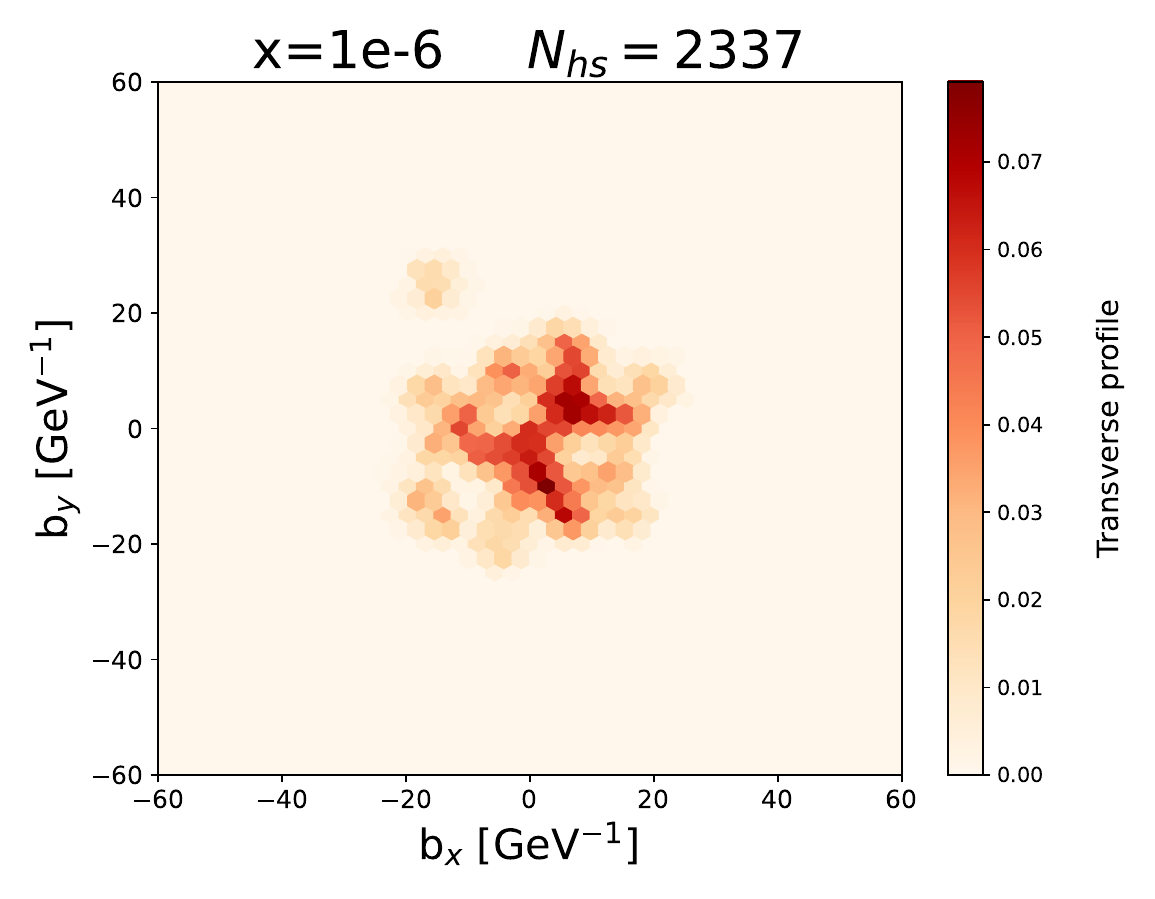}
\includegraphics[width=0.42\textwidth]{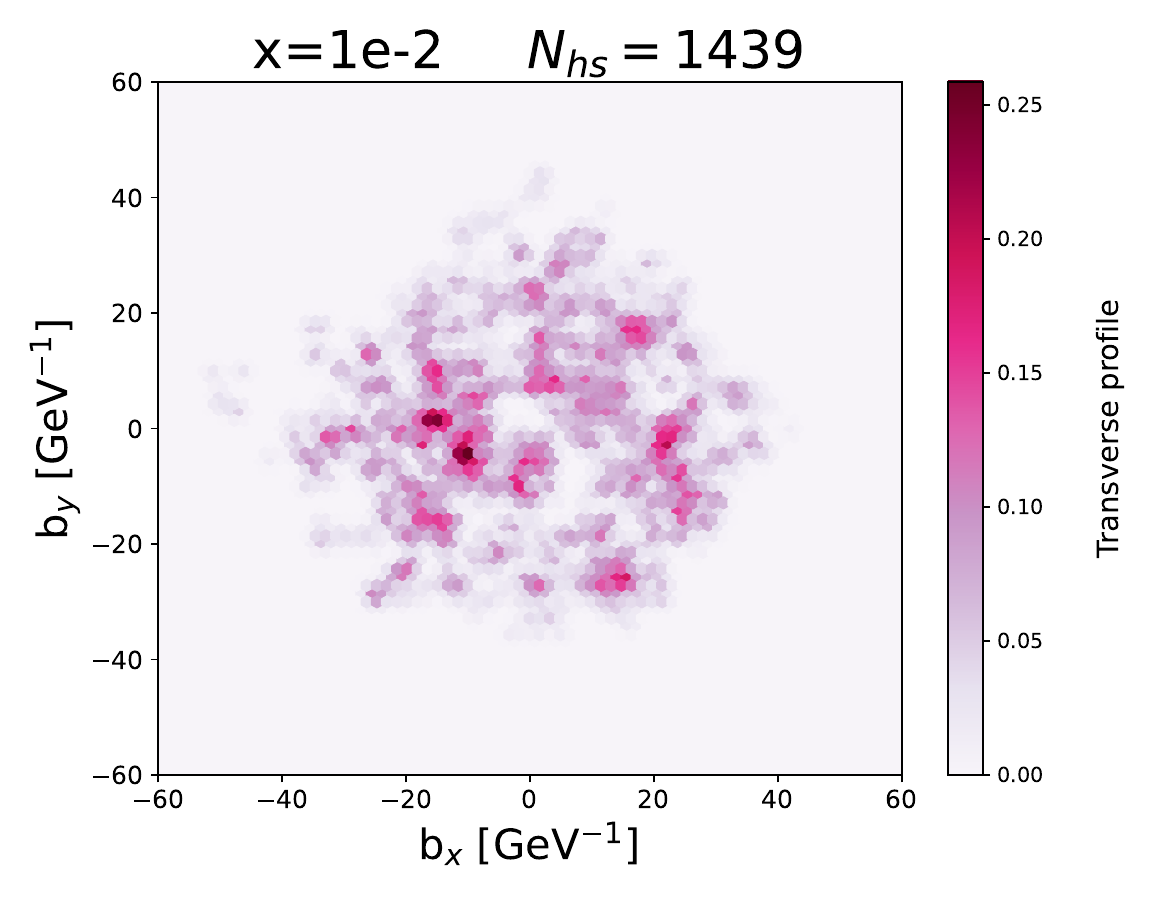}
\includegraphics[width=0.42\textwidth]{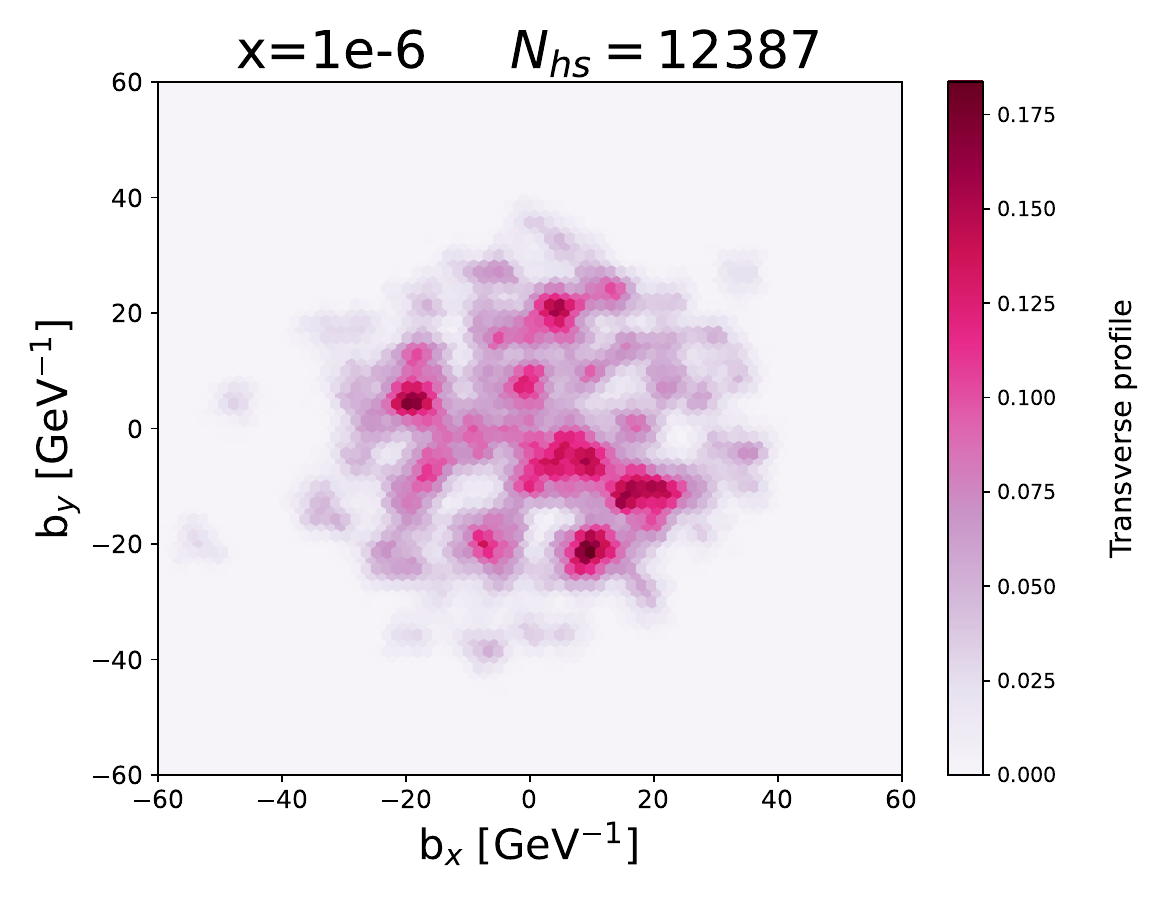}
 \centering
 \caption{Transverse profile of Ca (upper plots) and Pb (lower plots) generated using the energy - dependent hot - spot model at $ x=10^{-2}$ (left panels) and $ x=10^{-6}$ (right panels).}
 \label{fig:HS_nuclei}
\end{figure}

In the case of a nuclear target, the dipole - nucleus cross - section will be described using the Glauber - Gribov model \cite{glauber,gribov,mueller, Armesto:2002ny}, which implies
\begin{equation}
\frac{\dfr\sigma^{\rm dip}}{\dfr\rb_A} = 2\left[1-\left(1-\frac{1}{2A}\sigma_0 N(\tilde{x},\rr)T_{\rm A}(\rb_A)\right)^A\right]\,\,.
\label{dipole_lead}
\end{equation}
Moreover, the hot - spot model is applied to every individual nucleon and on top of that a random distribution of nucleons in the transverse plane according to the Woods - Saxon distribution \cite{DeVries:1987atn} is used  
\begin{equation}
T_{A} (\rb_h) = \frac{1}{2\pi B_{\rm hs}} \sum \limits_{i=1}^{A}
\frac{1}{N_{\rm hs}} \sum \limits_{j=1}^{N_{\rm hs}} 
 \exp \left( -\frac{\left( \rb_h - \rb_i- \rb_j\right)^2}{2B_{\rm hs}} \right).
\label{T_hs_Pb}
\end{equation}

In Figs. \ref{fig:HS_proton} and \ref{fig:HS_nuclei} we illustrate the evolution with the energy of the transverse proton and nuclear structure, respectively, showing example configurations generated within the energy - dependent hot - spot model. We have that as $x$ decreases, the increasing number of hot - spots leads to a denser and more homogeneous profile.

\section{Results}
\label{sec:results}

In this section we will investigate the impact of the sub - nucleon fluctuations on the total cross - section and $t$ - distributions considering the hot - spot model and the kinematical range that will be covered by the future electron - ion collider at BNL \cite{eic}. These two observables are directly related, since the total cross - section for the exclusive real photon production is given by
\begin{eqnarray}
\sigma^{\gamma^* h \rightarrow \gamma h (Y)} (W,Q^2) = \int_{-\infty}^0 dt \, \frac{d\sigma^{\gamma^{*}h \to {\gamma{h (Y)}}}}{dt}
\end{eqnarray}
where $W$ is the photon - hadron center - of - mass energy, $Q^2$ is the virtuality of the incoming photon and the differential cross - sections are  given by Eqs. (\ref{dsigdtcoherent}) and (\ref{dsigdtincoherent}) for coherent and incoherent interactions, respectively.

\begin{figure}[t]
	\begin{center}
   		\includegraphics[width=0.32\textwidth]{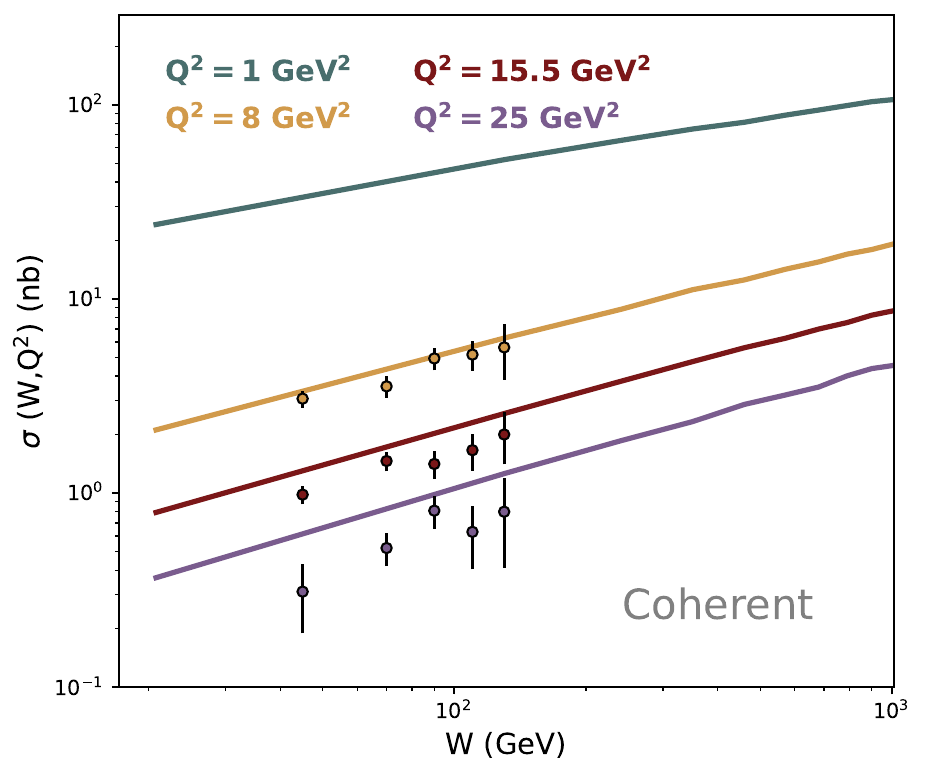}
		\includegraphics[width=0.32\textwidth]{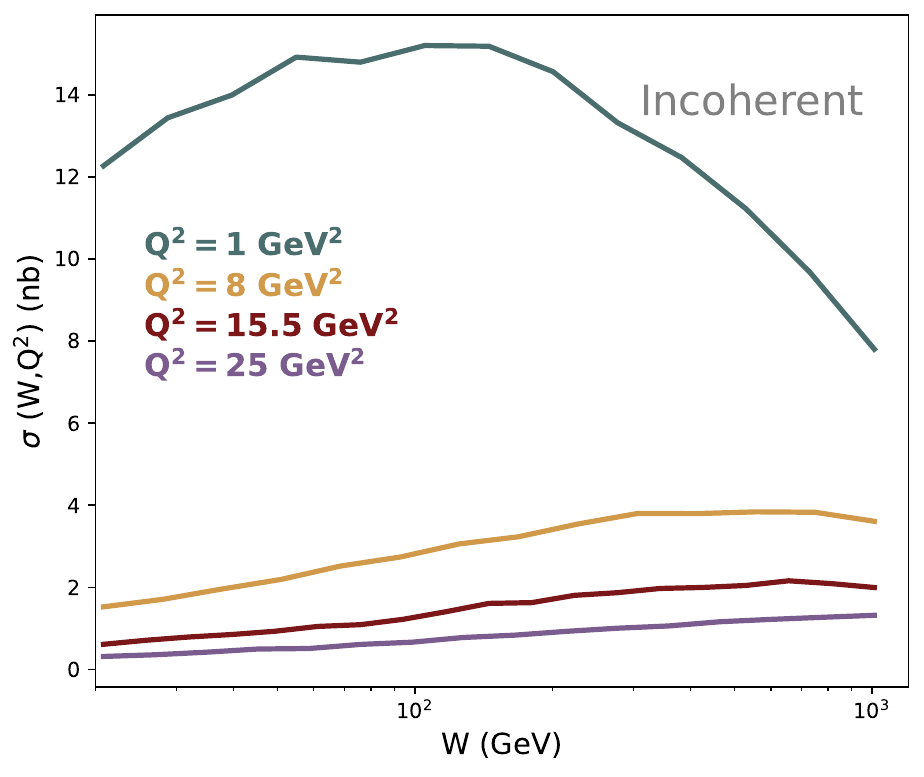}
        \includegraphics[width=0.32\textwidth]{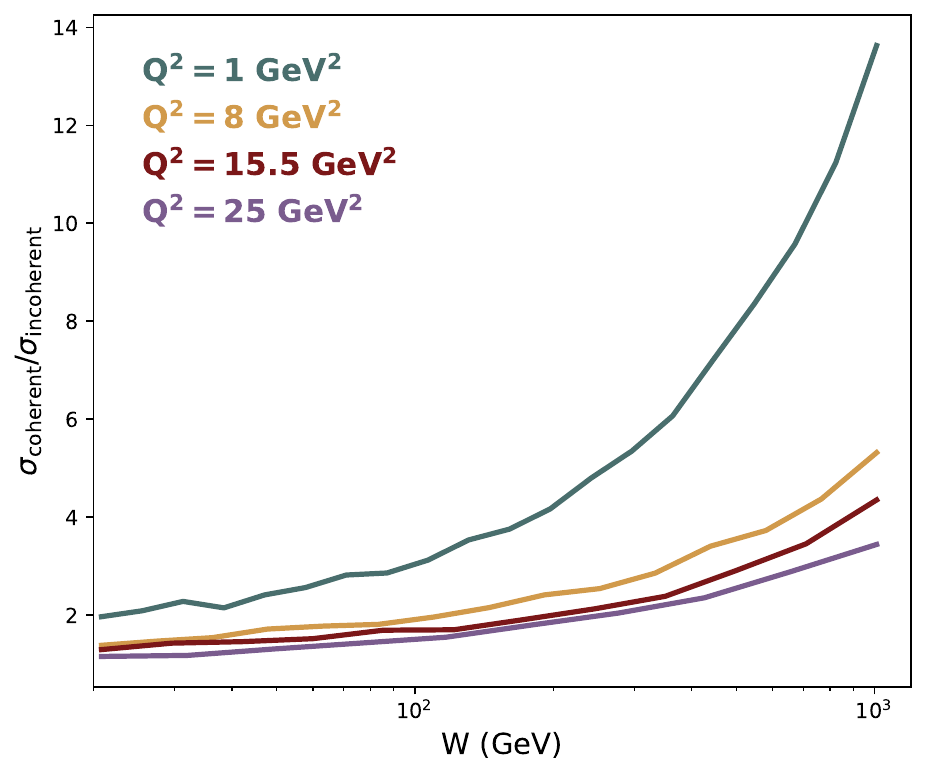}
 	\end{center}
        \caption{Predictions for the energy dependence of the cross - sections associated with coherent (left panel) and incoherent (center panel) interactions with a proton target. The results for the ratio between the coherent and incoherent cross - sections are presented in the right panel. Results for   different values of the photon virtuality $Q^2$. Data from H1 Collaboration \cite{data_dvcs}. }
   \label{fig:HS_proton_energy}
\end{figure}

Initially, in Fig. \ref{fig:HS_proton_energy}, we present the predictions for the energy dependence of the coherent (left panel) and incoherent (center panel)  cross - sections  assuming different values for the photon virtuality and a proton target. The data from H1 Collaboration \cite{data_dvcs} are presented for  comparison with our predictions. The cross - sections decrease for larger photon virtualities.  For coherent interactions, one has that the HERA data  are well described by the hot - spot model. In addition, we predict the presence of a turn - over in the energy dependence of the incoherent cross - section  for a proton target, as already predicted for vector mesons. However, in the case of the DVCS process,  the position of the maximum is strongly dependent of $Q^2$, occurring at smaller energies for lower photon virtualities.  The increase with the energy is also dependent of $Q^2$, with the slope decreasing for higher virtualities. In particular, for small virtualities, the changing in the energy dependence of the incoherent cross - section is predicted to occur in the energy range accessible in the EIC.  Finally, in the right panel of Fig. \ref{fig:HS_proton_energy}, we present our predictions for the energy dependence of the ratio between the coherent and incoherent cross - sections. We have that the ratio decreases with $Q^2$ and increases with the energy $W$. In particular, the presence of the turn - over in the incoherent cross - section implies a steeper increasing for $W \gtrsim  80$ GeV and low values of $Q^2$. In principle, such predictions could be tested in  future electron - ion colliders.

\begin{figure}[t]
	\begin{center}
       \includegraphics[width=0.45\textwidth]{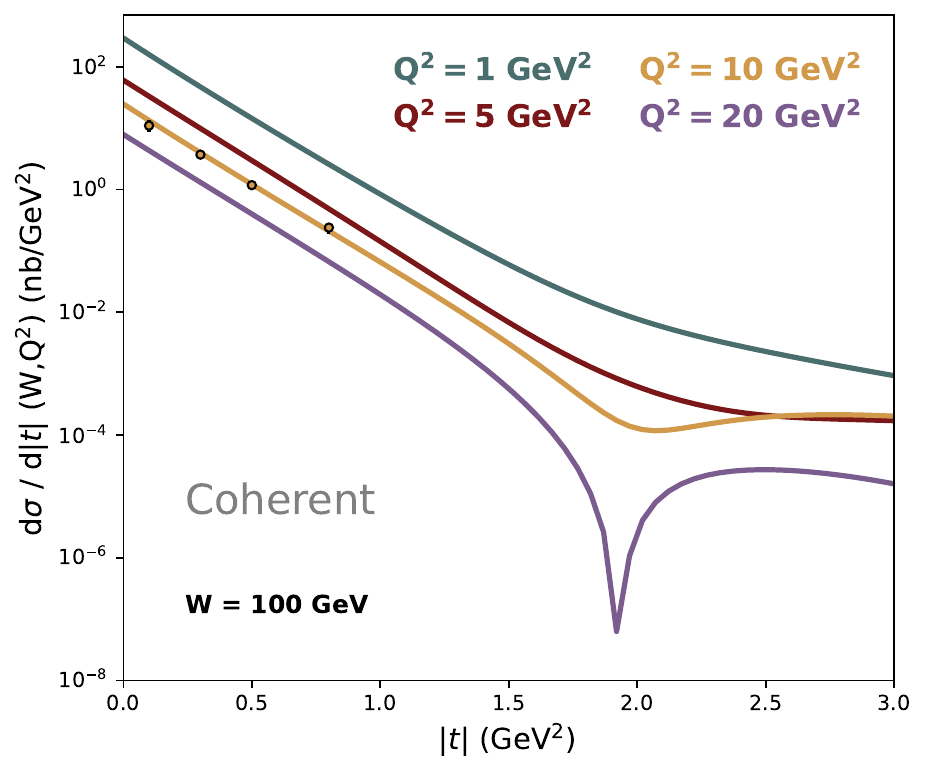}
        \includegraphics[width=0.45\textwidth]{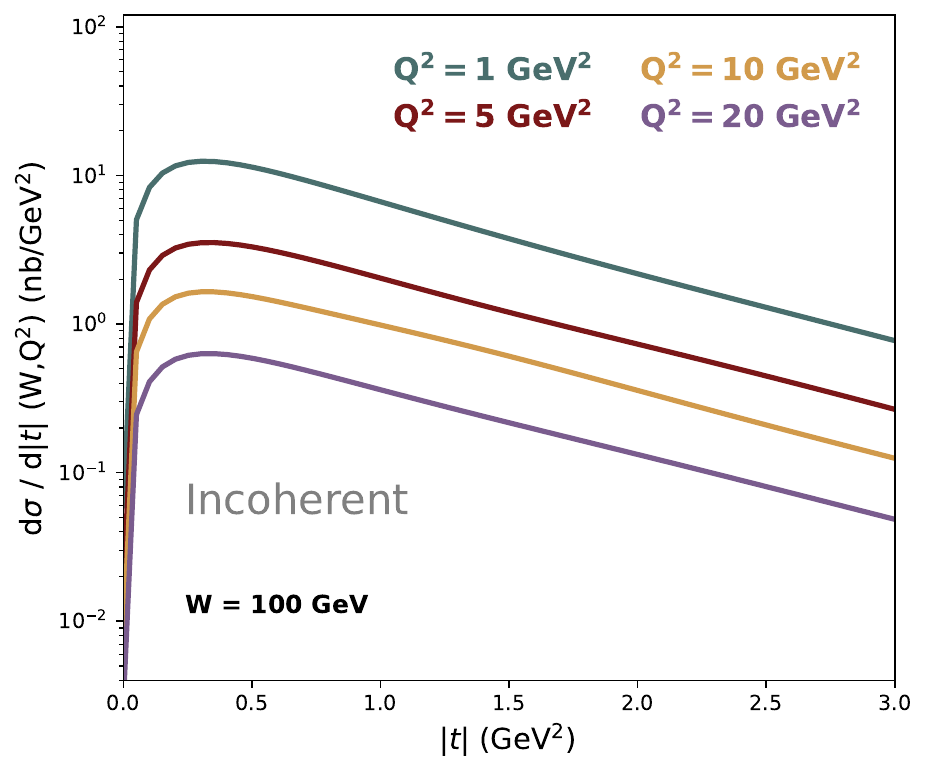}
 	\end{center}
        \caption{Predictions for the  differential $|t|$ - distributions   for the DVCS process associated with coherent (left panel) and incoherent (right panel) interactions with a proton target. Results for   different values of the photon virtuality $Q^2$. Data from H1 Collaboration \cite{data_dvcs}.}
   \label{fig:HS_proton_tdist}
\end{figure}

In Fig. \ref{fig:HS_proton_tdist}, we present the predictions for the  differential $|t|$ - distributions associated with the coherent (left panel) and incoherent (right panel)  cross - sections  assuming different values for the photon virtuality and a proton target. The data from H1 Collaboration \cite{data_dvcs} are presented for  comparison with our predictions. One has that the HERA data  are well described by the hot - spot model. In addition, we predict that the presence (or not) of a dip in the $t$ - distribution is dependent of the photon virtuality. In contrast, for incoherent interactions, a dip is not predicted, and the differential distribution vanishes when $|t| \rightarrow 0$, which is expected in the Good - Walker approach \cite{Good:1960ba}.

\begin{figure}[t]
	\begin{center}
		\includegraphics[width=0.32\textwidth]{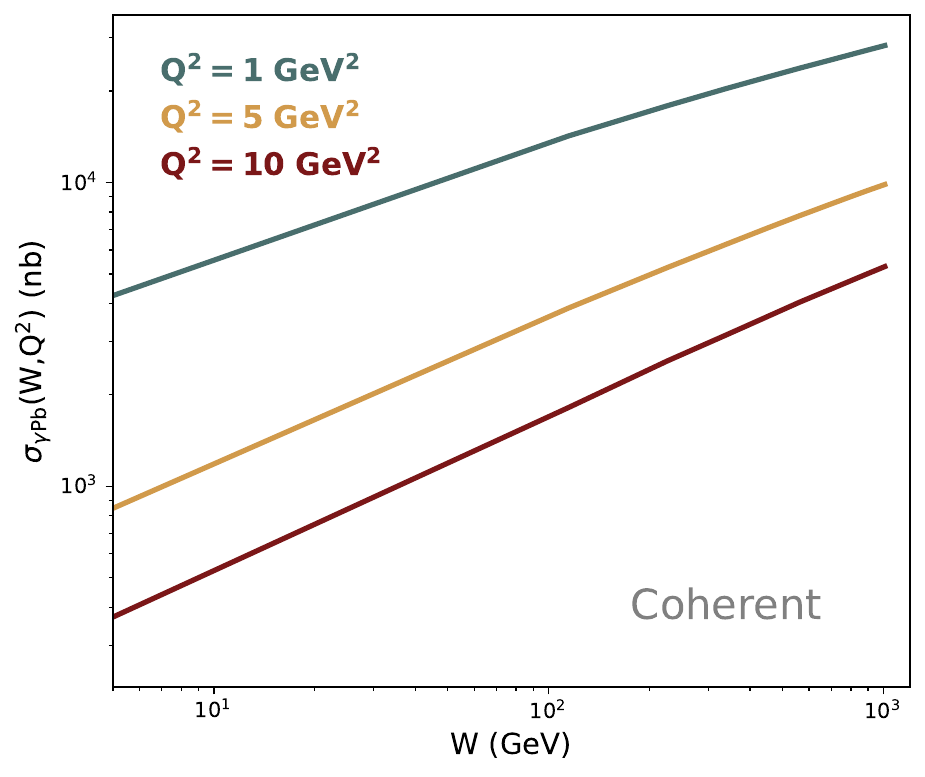}
		\includegraphics[width=0.32\textwidth]{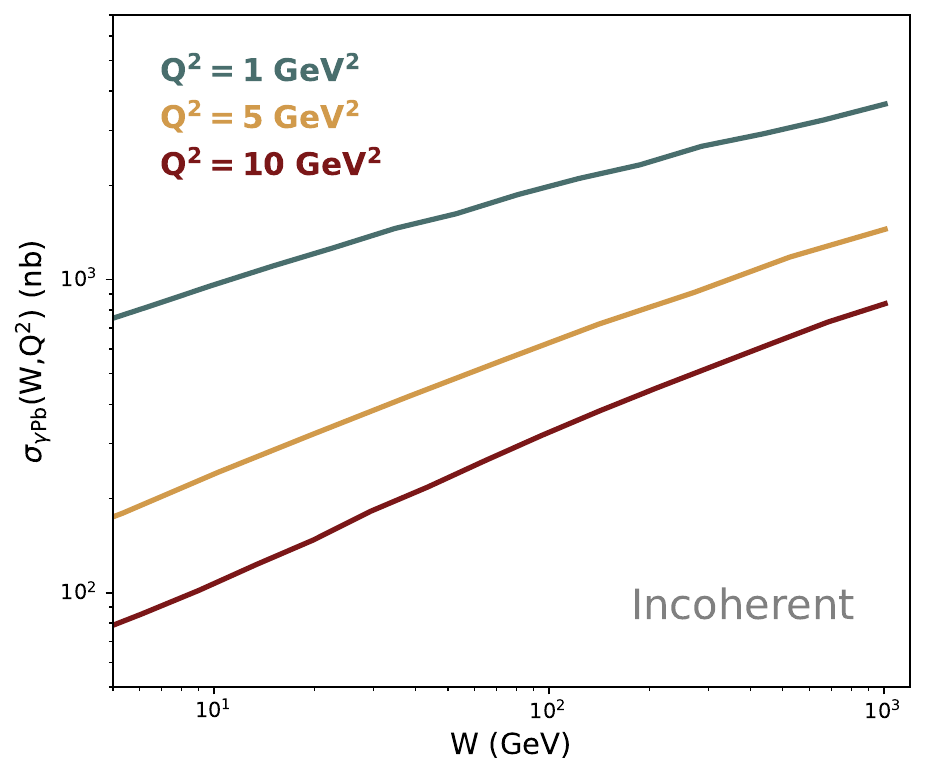}
        \includegraphics[width=0.32\textwidth]{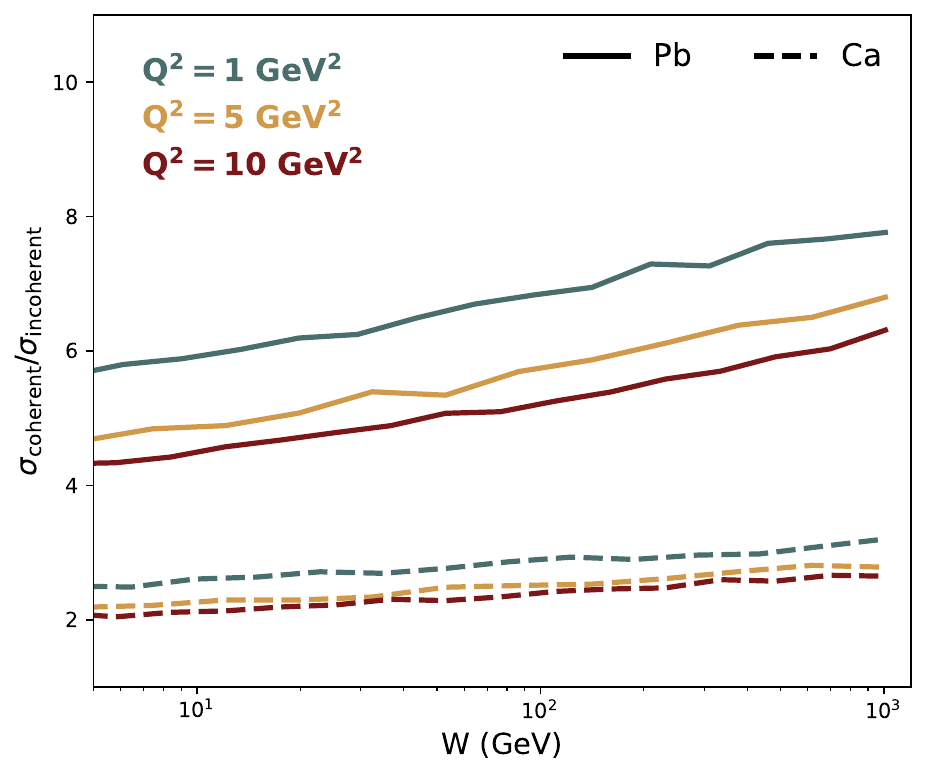}
	\end{center}
        \caption{Energy dependence of the total DVCS cross - section associated with coherent (left panel) and incoherent (center panel) interactions on Pb. Predictions for the ratio between the coherent and incoherent cross - sections are presented in the right panel, derived considering two distinct nuclear targets. Results  for different values of $Q^2$.}
    \label{fig:HS_nuclei_energy}
\end{figure}

In what follows, we will consider the DVCS process in a nuclear target. For completeness of our analysis, we will present our predictions for a heavy (Pb, $A = 208$) and a light (Ca, $A =  40$) nucleus. In Fig. \ref{fig:HS_nuclei_energy}  we present our predictions for the energy dependence of the coherent (left) and incoherent (center panel) cross - sections for the case of a lead target. As in the proton case, the coherent cross - section is larger than the incoherent one, with both decreasing with the photon virtuality. However, in contrast with the results for a proton target, a turn - over in the energy dependence of the incoherent cross - section is not predicted in the energy range considered. We have verified that similar results are also derived for $A = 40$.  The absence of a turn - over implies that the slope of the ratio between the coherent and incoherent cross - sections is not modified with the increasing of $W$, as observed in the right panel of Fig. \ref{fig:HS_nuclei_energy}. It is important to emphasize that our results indicate that the magnitude of the ratio and its energy dependence are dependent of $A$, with the ratio increasing with the atomic number and having a steeper slope for a heavier nucleus.

\begin{figure}[t]
	\begin{center}
		\includegraphics[width=0.45\textwidth]{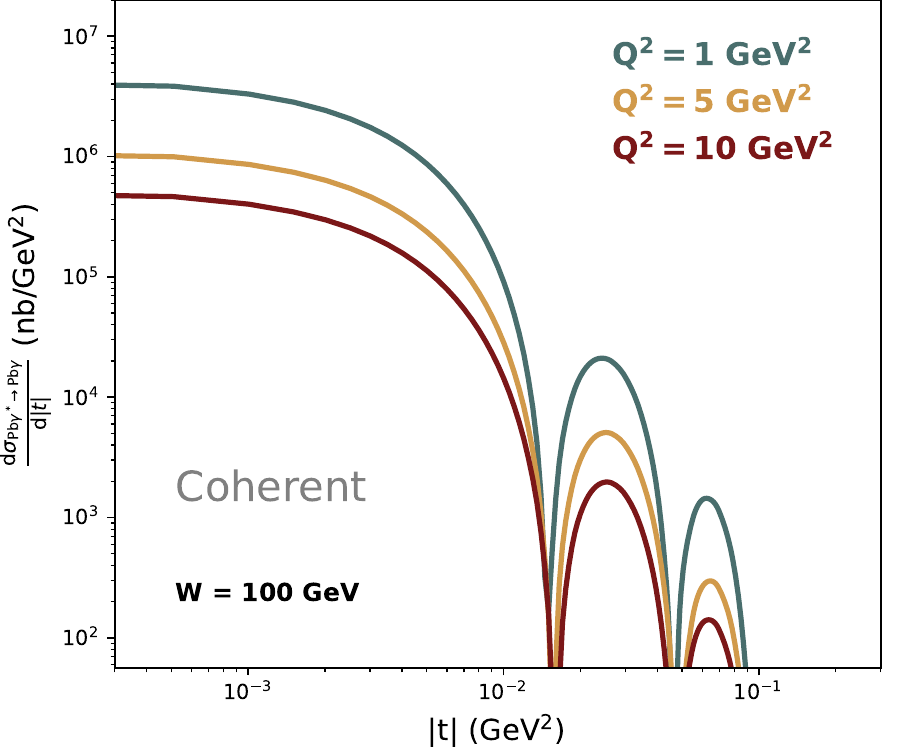}
		\includegraphics[width=0.45\textwidth]{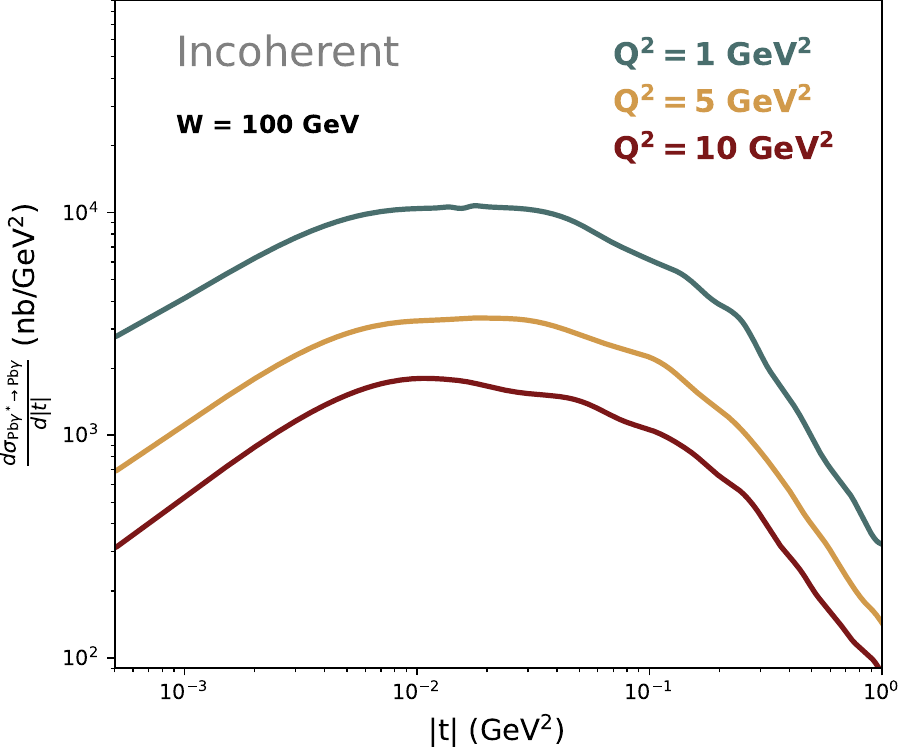}
		\includegraphics[width=0.45\textwidth]{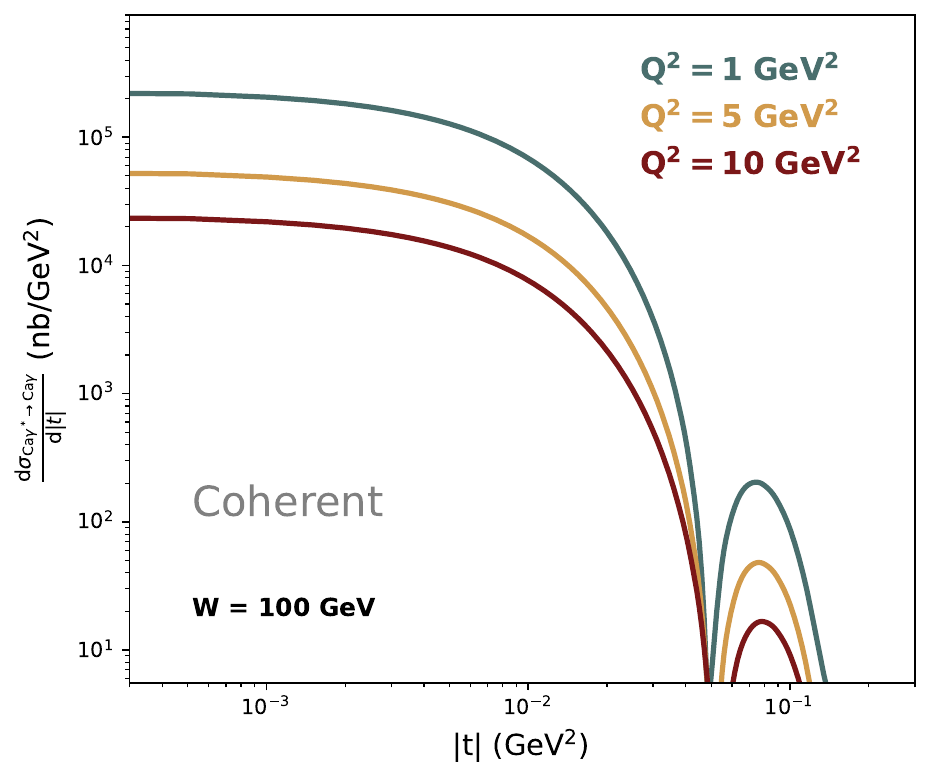}
		\includegraphics[width=0.45\textwidth]{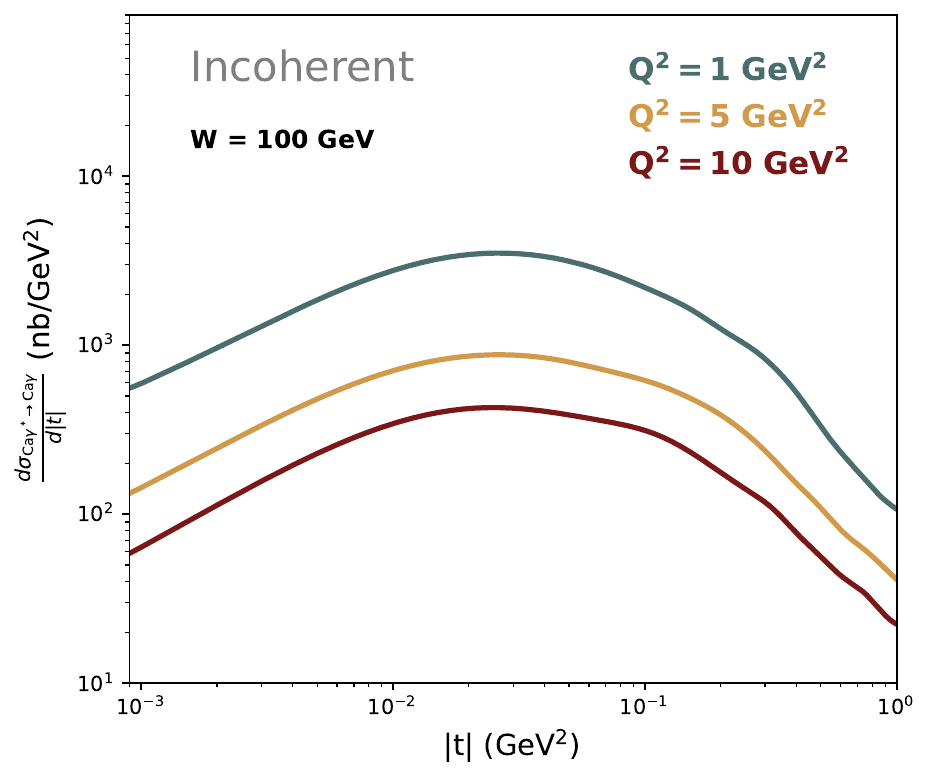}
	\end{center}
        \caption{Predictions for the differential $|t|$ - distributions for the DVCS process associated with  coherent (left panels) and incoherent (right panels) interactions with a $Pb$ (upper panels) and $Ca$ (lower panels) targets. Results for different values of  $Q^2$.}
         \label{fig:HS_nuclei_tdist}
\end{figure}

Finally, in Fig. \ref{fig:HS_nuclei_tdist} we present our predictions for the differential $|t|$ - distributions for the DVCS process associated with the coherent (left panels) and incoherent (right panels) interactions with a nuclear target. In the upper (lower) panels we present the results for $A = 208$ (40). The distributions decrease with the photon virtuality, as in the proton case. The coherent cross - section clearly exhibits the typical diffractive pattern. Dips are predicted for coherent interactions, with the positions being almost independent of photon virtuality. However, the position of the first dip is dependent of the atomic number, occurring at smaller values for a heavier nucleus. For incoherent interactions (right panels), the hot - spot model predicts that the cross - section decreases for $|t| \rightarrow 0$ and for $|t| \gtrsim 0.1$ GeV$^2$. Our results indicate that the position of the maximum is dependent on the photon virtuality and atomic number, occurring at larger values with the decreasing of $Q^2$ and for a heavier nucleus.  The comparison between these predictions and the future experimental data at the EIC will be an important check of the hot - spot model.

\section{Summary}
\label{sec:summary}
In this paper, we have investigated the coherent and incoherent DVCS processes assuming the energy - dependent hot - spot model to describe the sub - nucleon fluctuations. Predictions for the dependencies of the cross - sections on the energy, transverse momentum, photon virtuality and atomic number have been presented. Our results demonstrated that the hot - spot model predicts a turn - over in the energy dependence of the incoherent cross - section, with the position of the maximum being dependent of the photon virtuality. Moreover, we also predict a maximum in the $t$ - distribution for the nuclear incoherent cross - section, with the peak being dependent on $Q^2$ and $A$. Such predictions, directly associated with the hot - spot model, could be tested in the future electron - ion collider at BNL.

\begin{acknowledgments}
VPG would like to thank the members of the Czech Technical University in Prague for the warm hospitality during the beginning of this project. VPG was partially financed by the Brazilian funding agencies CNPq, FAPERGS and INCT-FNA  (Process No. 408419/2024-5). This work was partially funded by the Czech Science Foundation (GAČR), project No. 22-27262S.
\end{acknowledgments}


\begin{thebibliography}{99}


\bibitem{McDermott:2001pt}
M.~McDermott, R.~Sandapen and G.~Shaw,
Eur. Phys. J. C \textbf{22}, 655-666 (2002)

\bibitem{Favart:2003cu}
L.~Favart and M.~V.~T.~Machado,
Eur. Phys. J. C \textbf{29}, 365-371 (2003)

\bibitem{Kowalski:2006hc}
H.~Kowalski, L.~Motyka and G.~Watt,
Phys.\ Rev.\ D {\bf 74}, 074016 (2006).


\bibitem{Machado:2008tp}
M.~V.~T.~Machado,
Eur. Phys. J. C \textbf{59}, 769-776 (2009)

\bibitem{Aschenauer:2013hhw}
E.~C.~Aschenauer, S.~Fazio, K.~Kumericki and D.~Mueller,
JHEP \textbf{09}, 093 (2013)

\bibitem{Goncalves:2015goa}
V.~P.~Gon\c{c}alves and D.~S.~Pires,
Phys. Rev. C \textbf{91}, no.5, 055207 (2015)


\bibitem{Hatta:2017cte}
Y.~Hatta, B.~W.~Xiao and F.~Yuan,
Phys. Rev. D \textbf{95}, no.11, 114026 (2017)


\bibitem{Mantysaari:2020lhf}
H.~M\"antysaari, K.~Roy, F.~Salazar and B.~Schenke,
Phys. Rev. D \textbf{103}, no.9, 094026 (2021)

\bibitem{Bendova:2022xhw}
D.~Bendova, J.~Cepila, V.~P.~Gon{\c{c}}alves and C.~R.~Sena,
Eur. Phys. J. C \textbf{82}, no.2, 99 (2022)



\bibitem{Goncalves:2022wzq}
V.~P.~Gon{\c{c}}alves, D.~E.~Martins and C.~R.~Sena,
Eur. Phys. J. A \textbf{58}, no.2, 18 (2022)

\bibitem{Xie:2022sjm}
Y.~P.~Xie and V.~P.~Goncalves,
Phys. Rev. D \textbf{105}, no.1, 014033 (2022)

\bibitem{Xiang:2024fee}
W.~Xiang, D.~Cao and D.~Zhou,
Chin. Phys. C \textbf{48}, no.5, 054103 (2024)


\bibitem{Klein:2019qfb}
S.~R.~Klein and H.~M{\"a}ntysaari,
Nature Rev. Phys. \textbf{1}, no.11, 662-674 (2019)


\bibitem{Mantysaari:2020axf}
H.~M{\"a}ntysaari,
Rept. Prog. Phys. \textbf{83}, no.8, 082201 (2020)


\bibitem{Good:1960ba}
M.~L.~Good and W.~D.~Walker,
Phys. Rev. \textbf{120}, 1857-1860 (1960)



\bibitem{eic}
D.~Boer, M.~Diehl, R.~Milner, R.~Venugopalan, W.~Vogelsang, D.~Kaplan, H.~Montgomery, and S.~Vigdor {\it et al.},
arXiv:1108.1713 [nucl-th];\\
A.~Accardi, J.~L.~Albacete, M.~Anselmino, N.~Armesto, E.~C.~Aschenauer, A.~Bacchetta, D.~Boer, and W.~Brooks {\it et al.},
Eur.\ Phys.\ J.\ A {\bf 52}, no. 9, 268 (2016);\\  E.~C.~Aschenauer {\it et al.},
Rept.\ Prog.\ Phys.\  {\bf 82}, no. 2, 024301 (2019).

\bibitem{Cepila:2016uku}
J.~Cepila, J.~G.~Contreras and J.~D.~Tapia Takaki,
Phys. Lett. B \textbf{766}, 186-191 (2017)

\bibitem{Cepila:2017nef}
J.~Cepila, J.~G.~Contreras and M.~Krelina,
Phys. Rev. C \textbf{97}, no.2, 024901 (2018)



\bibitem{Cepila:2018zky}
J.~Cepila, J.~G.~Contreras, M.~Krelina and J.~D.~Tapia Takaki,
Nucl. Phys. B \textbf{934}, 330-340 (2018)

\bibitem{Cepila:2023dxn}
J.~Cepila, J.~G.~Contreras, M.~Matas and A.~Ridzikova,
Phys. Lett. B \textbf{852}, 138613 (2024)

\bibitem{Cepila:2025exl}
J.~Cepila, J.~G.~Contreras, M.~Matas and A.~Ridzikova,
Phys. Rev. C \textbf{113}, no.2, 2 (2026)


\bibitem{Bendova:2018bbb}
D.~Bendova, J.~Cepila and J.~G.~Contreras,
Phys. Rev. D \textbf{99}, no.3, 034025 (2019)


\bibitem{Krelina:2019gee}
M.~Krelina, V.~P.~Goncalves and J.~Cepila,
Nucl. Phys. A \textbf{989}, 187-200 (2019)

\bibitem{Nikolaev:1990ja}
N.~N.~Nikolaev and B.~G.~Zakharov,
Z. Phys. C \textbf{49}, 607-618 (1991)


\bibitem{Nikolaev:1991et}
N.~Nikolaev and B.~G.~Zakharov,
Z. Phys. C \textbf{53}, 331-346 (1992)

\bibitem{Morreale:2021pnn}
A.~Morreale and F.~Salazar,
Universe \textbf{7}, no.8, 312 (2021)

\bibitem{Golec-Biernat:1998zce}
K.~J.~Golec-Biernat and M.~Wusthoff,
Phys. Rev. D \textbf{59}, 014017 (1998)



\bibitem{glauber}
R. J. Glauber, in Lecture in Theoretical Physics, Vol. 1, edited by W. E. Brittin, L. G. Duham (Interscience, New York, 1959).


\bibitem{gribov}
V.~N.~Gribov,
Sov.\ Phys.\ JETP {\bf 29}, 483 (1969); Sov.\ Phys.\ JETP {\bf 30}, 709 (1970).


\bibitem{mueller}
A.~H.~Mueller,
Nucl.\ Phys.\ B {\bf 335}, 115 (1990).

\bibitem{Armesto:2002ny}
N.~Armesto,
Eur.\ Phys.\ J.\ C\ {\bf 26}, 35 (2013).
                           
\bibitem{DeVries:1987atn}
H.~De Vries, C.~W.~De Jager and C.~De Vries,
Atom. Data Nucl. Data Tabl. \textbf{36}, 495-536 (1987)


\bibitem{data_dvcs}
F.~D.~Aaron \textit{et al.} [H1],
Phys. Lett. B \textbf{681}, 391-399 (2009)

\end{thebibliography}

\end{document}